\documentclass[]{aastex631}
\usepackage{amsmath,comment}

\begin{document}
\title{Wide-Spectral-Band {Nuller} Insensitive to Finite Stellar Angular Diameter\\ with a One-Dimensional Diffraction-Limited Coronagraph}

\author[0000-0003-2690-7092]{Satoshi Itoh}
\author{Taro Matsuo}
\affiliation{Department of Science, Graduate School of Science, Nagoya University, Furocho, Chikusa, Nagoya, Aichi, 466-8602, Japan}
\author{Motohide Tamura}
\affiliation{Department of Astronomy, Graduate School of Science, The University of Tokyo, 7-3-1 Hongo, Bunkyo-ku, Tokyo 113-0033}
\affiliation{Astrobiology Center, 2-21-1 Osawa, Mitaka, Tokyo 181-8588, Japan}
\affiliation{National Astronomical Observatory of Japan, 2-21-1 Osawa, Mitaka, Tokyo 181-8588, Japan}

\begin{abstract}
{
Potentially habitable planets around nearby stars less massive than solar-type stars could join targets of the spectroscopy of the planetary reflected light with future space telescopes. {However, the orbits of most of these planets occur near the diffraction limit for 6-m diameter telescopes.} Thus, while securing contrast-mitigation ability under a broad spectral bandwidth and a finite stellar angular diameter, {we must maintain planetary throughput} even at the diffraction-limited angles to be able to reduce the effect of the photon noise within a reasonable observation time.} A one-dimensional diffraction-limited coronagraph (1DDLC) observes planets near the diffraction limit with undistorted point spread functions but has a finite-stellar diameter problem in wideband use. This study presents a method for wide-spectral-band nulling insensitive to stellar-angular-diameter by adding a fiber nulling with a Lyot-plane phase mask to the 1DDLC. Designing the pattern of the Lyot-plane mask function focuses on the parity of the amplitude spread function of light. Our numerical simulation shows that the planetary throughput (including the fiber-coupling efficiency) can reach about 11\% for about 1.35-$\lambda/D$ planetary separation almost independently of the spectral bandwidth. The simulation also shows the raw contrast of about $4\times10^{-8}$ (the spectral bandwidth of 25\%) and $5\times10^{-10}$ (the spectral bandwidth of 10\%) for $3\times 10^{-2}$ $\lambda/D$ stellar angular diameter. The planetary throughput depends on the planetary azimuthal angle, which may degrade the exploration efficiency compared to an isotropic throughput but {is partially offset} by the wide spectral band.
\end{abstract}


\section{Introduction} \label{sec:intro}
{Stellar coronagraphy} \citep{1939MNRAS..99..580L,1992A&A...261..499S}, a method for directly detecting light from exoplanets, can measure the wavelength dependency (spectra) of the light intensity reflected on the exoplanet. 
The spectra of light from exoplanets contain information on the planets' surface environment, including vegetation and atmosphere compositions \citep{2007ApJ...658..598K}.
Thus, obtaining the wavelength spectra of light from exoplanets may lead to the detection of a signature of extraterrestrial life \citep{2016JATIS...2d1212R}. 

Stellar coronagraphs \citep[e.g.,][]{1997PASP..109..815R,2001PASP..113..436B,2002ApJ...570..900K,2004EAS....12..185B,2005ApJ...630..631O,2005ApJ...633.1191M,2005OptL...30.3308F,2006dies.conf..485R,2008PASP..120.1112M,2009OExpr..1720515C,2017OExpr..25.7273B} mitigate the side-lobe of the stellar point spread functions (PSFs) to observe exoplanets with a moderate signal-to-noise ratio \citep{1994ApJ...425..348N} with the help of a speckle-nulling instrument \citep{2003ESASP.539..469K,2004SPIE.5487.1246L,2006ApJ...638..488B,2007CRPhy...8..365M,2009ApJ...693...75G}. 
While the raw contrast ratio is about $10^{-10}$ at the visible or near-infrared wavelength region in the case of Earth-Sun-like systems, potentially habitable planets around M-dwarves typically have a contrast of $10^{-7}$--$10^{-8}$ in the same wavelength region \citep{2009astro2010S.151K}. 
When we assume a telescope diameter $D$ of 6 m and observation wavelength $\lambda$ of 700 nm, the diffraction-limit scale $\lambda/D$ is about 0.024'', which approximately corresponds to a quarter times the separation angles of 10-pc-distant Sun-Earth analogs.   
Since potentially habitable planets around K-, M-dwarves have smaller orbital radii compared to ones around Solar-type stars, potentially habitable planets around K-, M-dwarves can have separation angles near the diffraction-limit scale $1\ \lambda/D$. 
Hence, we require small inner working angles of the high-contrast imaging instrument to observe these planets near the diffraction-limit scale $1\ \lambda/D$.
In addition, since the planets are extraordinarily dim, coronagraphs need high off-axis throughput near the diffraction-limit separation angles.
The observations also require the contrast mitigation ability that works with stellar non-zero angular diameters (e.g., about 0.00093'' for 10-pc-distant Sun analog) and non-zero spectral bandwidths (required for spectroscopic observation).
Furthermore, the contrast mitigation performance must remain stable over the whole integration time \citep{2019SPIE11117E..03P,2022JATIS...8c4001J}.

One-dimensional diffraction-limited coronagraphs \citep[1DDLCs;][]{Itoh+2020,2022SPIE12188E..4DI,2023PASP..135f4502I} {are promising candidates for observing planets} near the diffraction limit with sufficient throughput and distortion-free point spread functions. 
However, it has the problem that it can work in only narrow-band use (e.g., $\Delta \lambda/ \lambda = 0.3\%$ in the case of $10^{-10}$ of the contrast). 
To solve this problem, we investigated concepts of spectroscopic coronagraph \citep{Matsuo+2021,2022SPIE12188E..51O} and {a method \citep{2022AJ....163..279I,2022SPIE12188E..4DI} that consists of multiple successive uses of a modified version of the 1DDLC}. 
However, we have to implement complex instruments for these concepts.

{The spectroscopic coronagraph concept} uses at least two grating spectrometers.
{These coronagraphs addressed the problem of broadband use, but have the following limitations.}
One grating makes a one-dimensional spectrum of the host star's wavelength-dispersed diffraction image on the focal plane.
Another grating serves as the reverse element for the previous one concerning the wavelength dispersion and secures white light pupil after the image plane where the spectrum emerges. 
In addition, in the spectroscopic coronagraph, we have to slightly modify the 1DDLC mask pattern through a projective transformation \citep{Born_Wolf_2019} to fit the scales of the dispersed diffraction images of the different wavelengths approximately.
Furthermore, a single spectroscopic coronagraph cannot sufficiently mitigate the contrast between the potentially habitable planets and their host stars due to the non-zero stellar diameters.
 We require the serial use of at least two sets of spectroscopic coronagraphs to resolve the problem.

{The concept of multiple serial uses of a modified version of the 1DDLC comes from the fact that due to the difference between design and actual wavelengths, the 1DDLC outputs stellar leak with a flat complex amplitude (phase and amplitude) inside the Lyot stop.
The leak amplitude on the Lyot pupil serves as an aberration-free input for another following coronagraph system.  
Hence, serial use of the 1DDLC gains immunity to the wide-spectral-band use.
However, this concept reduces the instrumental throughput by increasing the number of serial uses because the single 1DDLC has an off-axis throughput of about 50\%. 
A previous study \citep{2022AJ....163..279I} showed that we can mitigate the degradation in the throughput due to the serial uses by modifying the patterns of the focal-plane mask and Lyot stop of the 1DDLC to complex patterns compared to the original ones for the 1DDLC.}

In this study, we present a new and simple method to achieve a wideband nulling with insensitivity to stellar angular diameters by adding a fiber nulling \citep{2018ApJ...867..143R,2020AJ....160..210W} with a kind of Lyot-plane mask \citep{2015A&A...583A..81R} to the 1DDLC.
To determine the mask profile of the Lyot-plane mask, we focus on the parity of the amplitude spread function of the stellar leak.
This is because parity is a concept independent of the scales of the focal-plane amplitude spread function, in other words, the wavelength of light; focal-plane amplitude spread functions scale with the wavelength of light when they come from a common pupil function.
In Section 2, we theoretically describe the new method.
In Section 3, we perform a numerical simulation to assess the achievable performance of the new method.
Section 4 discusses our interpretation of the simulation results.
Section 5 summarizes the content of this study.

\section{Theory} \label{sec:theory}
{We show a brief review of the 1DDLC in Section \ref{suma} before explaining the new method; see \citet{Itoh+2020} for derivation of the 1DDLC.}
\subsection{Review of the One-Dimensional Diffraction-Limited Coronagraph\label{suma}}
We use the pupil coordinates $\vec{\alpha}=(\alpha,\beta)$ normalized by the telescope diameter $D$ and the focal coordinates $\vec{x}=(x,y)$ normalized by $\lambda/D$, where $\lambda$ donates the light wavelength to be observed: the normalization factors $D$ and $\lambda/D$ scale with the magnification factors of pupils and images.
The following function definitions work throughout this paper:
\begin{equation}
    \mathrm{rect}\left[v\right]=\begin{cases}
1 \ \ \left(\left|v\right|<\frac{1}{2}\right) \\
\frac{1}{2} \ \ \left(\left|v\right|=\frac{1}{2}\right) \\
0 \ \ \left(\left|v\right|>\frac{1}{2}\right) 
\end{cases}
\end{equation}
and
\begin{equation}
    \mathrm{sinc}\left(x\right)=\frac{\sin{\left(\pi x\right)}}{\pi x}.
\end{equation}

In the 1DDLC, the focal-plane mask modulates the amplitudes and phases of light so that the light from the on-axis source has no transmittance through the Lyot Stop and that light from off-axis sources transmits through the Lyot Stop.
The 1DDLC has the following pupil aperture function $P(\vec{\alpha})$, Lyot stop aperture function $L(\vec{\alpha})$, and focal-plane mask function $M(\vec{x})$:
\begin{equation}
P(\vec{\alpha})=L(\vec{\alpha})=\mathrm{rect}\left[\alpha\right]\mathrm{rect}\left[\beta\right]\label{e3}
\end{equation}
and
\begin{equation}
M(\vec{x})=a\left(1-2\mathrm{sinc}\left(2x\right)\right),\label{e4}
\end{equation}
where the constant factor $a$ takes $0.697...$ so that $\left|M(\vec{x})\right|\leq 1$.
We see how the coronagraph works using the following orthonormal complete base functions $\left\lbrace b_{kl}\left(\vec{\alpha}\right)\left|k,l \in \mathbb{Z} \right. \right\rbrace$ :
\begin{equation}
    b_{kl}\left(\vec{\alpha}\right)=P(\vec{\alpha})e^{2\pi i\left(k \alpha +l \beta\right)}.
\end{equation}
Expanding an arbitrary pupil amplitude $v(\vec{\alpha})$ on the pupil aperture with the base functions leads to the following expression:
\begin{equation}
    v\left(\vec{\alpha}\right)=\sum_{k=-\infty}^{\infty}\sum_{l=-\infty}^{\infty}w_{kl} b_{kl}\left(\vec{\alpha}\right),
\end{equation}
where
\begin{equation}
w_{kl}= \int_{-\infty}^{\infty}\!\!\!\! d\alpha \int_{-\infty}^{\infty}\!\!\!\! d\beta \ b_{kl}\left(\vec{\alpha}\right)^{\ast} v\left(\vec{\alpha}\right).
\end{equation}
The coronagraph defined by Equation (\ref{e3}) and (\ref{e4}) acts as the following linear transformation:
\begin{eqnarray}
w_{kl}  &\to& a \left(1-\delta_{k0}\right)w_{kl},
\end{eqnarray}
where the symbol $\delta_{kl}$ means Kronecker delta.
Hence, the coronagraph nulls the weights $\left\lbrace w_{0l}\left|l \in \mathbb{Z} \right. \right\rbrace$  and transmits the weights $\left\lbrace w_{kl}\left|k,l \in \mathbb{Z}, k\neq 0 \right. \right\rbrace$ multiplying the constant factor $a$ of the focal-plane mask.

We can express a tilt aberration of tiny-amount direction cosines of $(\Delta \theta_{x},\Delta \theta_{y})$ on the pupil plane as the following expression:
\begin{eqnarray}
    v_{\mathrm{tilt}}\left(\vec{\alpha}\right)&=&P(\vec{\alpha})e^{2\pi i\left(\Delta \theta_{x} \alpha +\Delta \theta_{y} \beta\right)}\label{a9}\\
    &\sim&b_{00}\left(\vec{\alpha}\right)+2\pi i P(\vec{\alpha})\left(\Delta \theta_{x} \alpha +\Delta \theta_{y} \beta\right).
\end{eqnarray}
The coronagraph defined by Equation (\ref{e3}) and (\ref{e4}) works to the input amplitude $ v_{\mathrm{tilt}}\left(\vec{\alpha}\right)$ as the following transformation:
\begin{eqnarray}
 v_{\mathrm{tilt}}\left(\vec{\alpha}\right)  &\to&  v_{\mathrm{tilt}}^{\mathrm{out}}\left(\vec{\alpha}\right)=2\pi i a P(\vec{\alpha})\Delta \theta_{x} \alpha.
\end{eqnarray}
The Fourier transform of the output amplitude on the Lyot stop yields focal amplitude $u(\vec{x})$ on the detector plane:
\begin{eqnarray}
u(\vec{x})= \int_{-\infty}^{\infty}\!\!\!\! d\alpha \int_{-\infty}^{\infty}\!\!\!\! d\beta \ v_{\mathrm{tilt}}^{\mathrm{out}}\left(\vec{\alpha}\right)e^{-2\pi i\left(x \alpha +y \beta\right)}.
\end{eqnarray}
The pupil amplitude $\ v_{\mathrm{tilt}}^{\mathrm{out}}\left(\vec{\alpha}\right) $ is an odd function about the coordinate $\alpha$ and an even function about the coordinate $\beta$.
Since the Fourier transform keeps parity (whether the function is even or odd) of operand functions, the focal amplitude $u(\vec{x})$ on the detector plane is also an odd function about the coordinate $x$ and an even function about the coordinate $y$:
\begin{eqnarray}
    u(-x,y)&=&-u(x,y),\nonumber \\
    u(x,-y)&=&u(x,y).
    \label{uu}
\end{eqnarray}
Since the pupil amplitude $\ v_{\mathrm{tilt}}^{\mathrm{out}}\left(\vec{\alpha}\right) $ is proportional to the magnitude of the tilt aberration $\Delta \theta_{x}$, the intensity of the leak on the detector plane is proportional to $\left(\Delta \theta_{x}\right)^2$.
Hence, the 1DDLC system shows the second-order sensitivity to the tilt aberration due to the non-zero diameter of the central stars.

When the observation wavelength $\lambda$ differs from the design wavelength $\lambda_{d}$ of the focal-plane mask, we must change the mask function from Equation (\ref{e4}) to the following:
\begin{equation}
M_{\Lambda}(\vec{x})=a\left(1-2\mathrm{sinc}\left(2\Lambda x\right)\right),
\end{equation}
where we defined that $\Lambda=\lambda/\lambda_{d}$.
Since the deviation $M_{\Lambda}(\vec{x})-M_{1}(\vec{x})$ of the mask function from the ideal case ($\Lambda=1$) is an even function about $x$ and $y$, the leak amplitude $v(x,y)$ caused by this deviation is also an even function about $x$ and $y$ on the focal plane after the Lyot stop:
\begin{eqnarray}
    v(-x,y)&=&v(x,y),\nonumber \\
    v(x,-y)&=&v(x,y).
    \label{vv}
\end{eqnarray}
\subsection{Overview}
\begin{figure}[htb]
    \centering
    \includegraphics[width=0.7\textwidth]{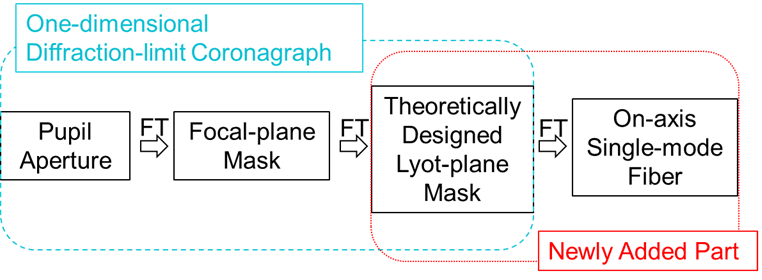}
    \caption{Schematics of the whole configuration of the new method. The abbreviation FT means ``Fourier Transform" of the Fraunhofer diffraction of light between focal and pupil planes. The parts encircled by the cyan dashed curve compose the 1DDLC. The red dotted curve encircled the newly added parts in the new method.  }
    \label{0.5}
\end{figure}

The new method consists of the 1DDLC (Appendix \ref{suma}) and newly added parts: a theoretically designed Lyot-plane mask and a single-mode fiber located at the on-axis point (Figure \ref{0.5}).
The 1DDLC transmits the stellar leaks (due to non-zero stellar angular diameter and spectral bandwidth) to its pupil plane where the Lyot stop is located (Lyot plane). 
We design the Lyot-plane mask so that it removes these leaks achromatically and delivers sufficient planetary light to spectrometers through the on-axis single-mode fiber.

\subsection{Parity of Leaks' Amplitude Spread Function}
We focus on the parity of the amplitude spread functions to design the appropriate Lyot-plane mask and build the new method.
{The Fraunhofer diffraction integral of a pupil function is a Fourier transform, which scales with the wavelength of light. }
This scaling by the wavelengths limits the bandwidths of phase-mask coronagraphy to narrow bands.
Parity is a concept independent of the scales of the focal-plane amplitude spread function, in other words, the wavelength of light, thus it is worth focusing on the parity of the amplitude spread functions to design a method less dependent on the wavelength of light.

We can write any amplitude spread function on pupil and focal planes as the sum of its four different components. 
The first component is a function that is an even function for the first and second arguments (we refer to this as EE component), the second is even for the first argument but is odd for the second arguments (EO component), the third is odd for the first and even for the second arguments (OE component), and the fourth is odd for both arguments (OO component). 
Formally, we can expand an arbitrary amplitude spread function $f(s,t)$ as a sum of the four components concerning their parities as follows:
\begin{equation}
    f(s,t)=f_{\mathrm{EE}}(s,t)+f_{\mathrm{OE}}(s,t)+f_{\mathrm{EO}}(s,t)+f_{\mathrm{OO}}(s,t),\label{eq1}
\end{equation}
where 
\begin{eqnarray}
    f_{\mathrm{EE}}(s,t)&=&\frac{f(s,t)+f(-s,t)+f(s,-t)+f(-s,-t)}{4},\\
    f_{\mathrm{OE}}(s,t)&=&\frac{f(s,t)-f(-s,t)+f(s,-t)-f(-s,-t)}{4},\\
    f_{\mathrm{EO}}(s,t)&=&\frac{f(s,t)+f(-s,t)-f(s,-t)-f(-s,-t)}{4},\\
    f_{\mathrm{OO}}(s,t)&=&\frac{f(s,t)-f(-s,t)-f(s,-t)+f(-s,-t)}{4},
\end{eqnarray}
and the coordinates $(s,t)$ represents the focal coordinates $(x,y)$ or the pupil coordinate $(\alpha,\beta)$.
The subscripts $\mathrm{p_1 p_2}$ in the notation of the above components $f_{\mathrm{p_1 p_2}}(s,t)$ ($\mathrm{p_1, p_2} \in \left\lbrace\mathrm{E,O}\right\rbrace$) express the parity (E: even, O: odd) about the first argument $s$ and the second arguments $t$, respectively. 
Since the Fourier transform does not change the parity of an arbitrary function, Fraunhofer diffraction between the focal and pupil plane preserves the parities of the light amplitudes.
We can achieve at least the fourth-order null by erasing the leak amplitude of the least-order term of tilt aberration; this least-order term of tilt aberration contains only the OE component (Section \ref{suma}).
The deviation of the actual wavelength from the design wavelength of the focal-plane mask creates only the EE component (Section \ref{suma}). 
\subsection{Parity Response of Single-Mode Fiber Located at Optical Axis}
In the new method, we use a single-mode fiber such that the fiber center matches the focal-plane origin $\vec{x}=\left(0,0\right)$; we assume that the stellar center matches the origin if no star-light-removing optical device including coronagraph exists. 
We assume that the transmission mode function $g(\vec{x})$ of the single-mode fiber is a function with only an EE component.
We can calculate the fiber-transmitted intensity $I$ using the following expression \citep{1982ApOpt..21.2671W}:
\begin{equation}
I=\left| \int_{-\infty}^{\infty}\!\!\!\! dx \int_{-\infty}^{\infty}\!\!\!\! dy \ g(x,y)f(x,y)\right|^2. \label{e6}
\end{equation}
Substituting Equation (\ref{eq1}) to Equation (\ref{e6}) leads to the intensity transmitted through the single-mode fiber for an arbitrary focal amplitude $f\left(\vec{x}\right)$ as follows.:
\begin{eqnarray}
    I&=&\left| \int_{-\infty}^{\infty}\!\!\!\! dx \int_{-\infty}^{\infty}\!\!\!\! dy \ g(x,y)\left(f_{\mathrm{EE}}(x,y)+f_{\mathrm{OE}}(x,y)+f_{\mathrm{EO}}(x,y)+f_{\mathrm{OO}}(x,y)\right)\right|^2 \nonumber\\
    &=&\left| \int_{-\infty}^{\infty}\!\!\!\! dx \int_{-\infty}^{\infty}\!\!\!\! dy \ g(x,y)f_{\mathrm{EE}}(x,y)\right|^2,\label{222}
\end{eqnarray}
where we used the odd-function nature of $f_{\mathrm{OE}}(x,y), f_{\mathrm{EO}}(x,y)$, and $f_{\mathrm{OO}}(x,y)$.
Equation (\ref{222}) shows that the amplitudes belonging to the OE, EO, and OO components fail to transmit the single-mode fiber located at the origin.
\subsection{Design of Lyot-Plane Phase Mask}
\begin{figure}[htb]
    \centering
    \includegraphics[width=\textwidth]{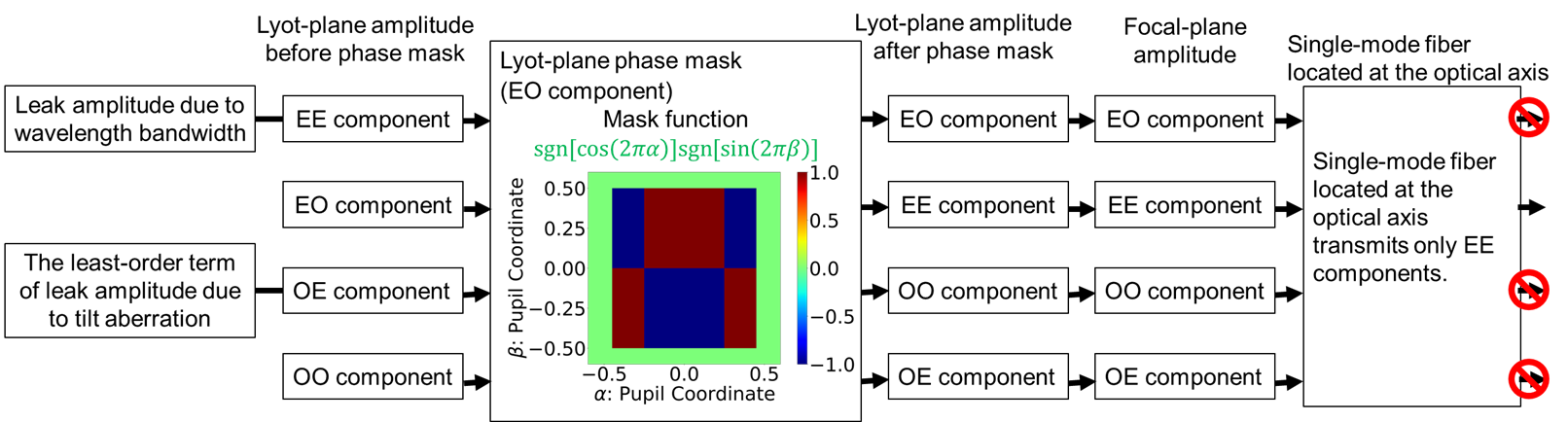}
    \caption{A block diagram showing how the new concept eliminates undesirable leakage amplitudes. See the main text for definitions of EE, EO, OE, and OO components. The red prohibited symbols mean that the associated components fail to transmit the single-mode fiber.}
    \label{1}
\end{figure}

To achieve the wideband fourth-order null by eliminating OE components (the leak from the least-order term of the tilt aberration) and EE components (the leak due to the wavelength deviation from the design wavelength), we need to change the OE  and EE components to components other than EE components.
Multiplying a function that has only an EO component to the functions of OE and EE components leads to the function of OO and EO components, respectively; this multiplication satisfies the above requirement.
Hence, the mask function of the Lyot-plane phase mask must include only an EO component to achieve the wideband fourth-order null.  We compiled how the new concept works to remove undesirable leak amplitudes as a block diagram in Figure \ref{1}.

\begin{figure}[htb]
    \centering
    \includegraphics[width=\textwidth]{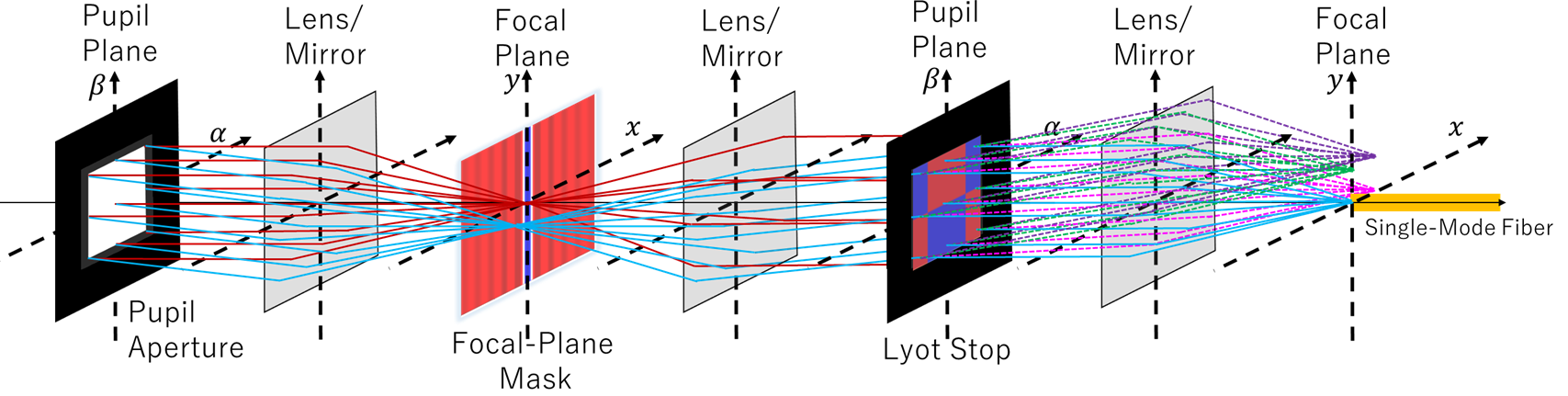}
    \caption{A schematic showing how planetary light can transmit the single-mode fiber located at the optical axis. The blue and red lines before the Lyot plane indicate off-axis planetary and on-axis steller light rays, respectively. The Lyot-stop phase-modulation mask (see also Figure \ref{1}) works as a two-dimensional diffraction grating, which creates multiple diffraction peaks on the next focal plane. We show the rays belonging to the four diffraction peaks with the diffraction orders $(1,1)$, $(-1,1)$, $(1,-1)$, and $(-1,-1)$ using different colors for every diffraction peak. When the planet has proper separation angles depending on mask parameters, one of the diffraction peaks exists near the optical axis and transmits the single-mode fiber effectively. }
    \label{fig:my_label1}
\end{figure}

The additional requirement for the Lyot-plane phase mask concerns the planetary throughput.
To secure the planetary throughput via the single-mode fiber located at the optical axis, we can use the following sinusoidal modulation pattern with only an EO component:
\begin{eqnarray}
    s^{\mathrm{sinu}}_{qr}\left(\vec{\alpha}\right)&=&\cos{\left(2\pi q \alpha\right)}\sin{\left(2\pi r \beta\right)} \nonumber \\
    &=&\frac{1}{4i}\left(e^{2 \pi i q \alpha}+e^{-2 \pi i q \alpha}\right)\left(e^{2 \pi i r \beta}+e^{-2 \pi i r \beta}\right),
\end{eqnarray}
where $q$ and $r$ are the design parameters of the mask.
The modulation pattern $ s^{\mathrm{sinu}}_{qr}\left(\vec{\alpha}\right)$ works as a two-dimensional diffraction grating with non-zero diffraction efficiency for only the diffraction orders of $(1,1)$, $(-1,1)$, $(1,-1)$, and $(-1,-1)$. 
The single peak of the focal amplitude exists at the optical axis and transmits the single-mode fiber efficiently (Figure \ref{fig:my_label1}).

Adopting the following rectangular-wave-like phase modulation pattern improves the planetary throughput and the easiness of manufacturing the mask compared to the case with the sinusoidal pattern:
\begin{eqnarray}
    s^{\mathrm{rect}}_{qr}\left(\vec{\alpha}\right)&=&\mathrm{sgn}\left[\cos{\left(2\pi q \alpha\right)}\right]\mathrm{sgn}\left[\sin{\left(2\pi r \beta\right)}\right],
   \label{eq9}
\end{eqnarray}
where the symbol $\mathrm{sgn}\left[...\right]$ means the sign function.
We define the sign function as follows:
\begin{equation}
    \mathrm{sgn}\left[z\right]=\begin{cases}
-1 \ \ \left(z<0\right) \\
0  \ \ \left(z=0\right) \\
1 \ \ \left(z>0\right).
\end{cases}
\end{equation}
Since the modulation pattern $s^{\mathrm{rect}}_{qr}\left(\vec{\alpha}\right)$ takes the values of only $-1$ or $1$ (i.e., $e^{\pi i}$ or $e^{0 i}$), the modulation includes only $\pi$-radian phase modulation, resulting in the high planetary throughput and the mask-manufacturing simplicity. See also Appendix \ref{ape2} concerning another expression of the Lyot-plane phase mask function.

\subsection{Other Instrumental Conditions}
For a realistic performance evaluation of the new concept, we have to apply the following instrumental conditions. 
(1)We use the 1DDLC summarized in Section \ref{suma}.
(2)We utilize the Lyot-plane phase-mask function shown in Equation (\ref{eq9}) with the following parameters: $(q,r)=(1,1)$.
Because we are most interested in the performance of the case with the innermost planetary angular separation from the perspective of complementarity to other coronagraphs, we adopt the above mask parameters.
(3)We assume the following transmission mode function of the single-mode fiber located at the optical axis of the final focal plane: $g(x,y)=\mathrm{sinc(x)sinc(y)}$.
The heart of this assumption is only the parity of the transmission mode function. We are not interested in the precise functional form of the transmission mode function here.

\section{Simulation} \label{sec:simulation}
\subsection{setup}
To evaluate the theoretically achievable performance of the new concept, we perform a numerical simulation with the following setup.  
The simulation includes the following parameters expressing different simulation cases.
\begin{enumerate} 
    \item The spectral bandwidth $\frac{\Delta\lambda}{\lambda_c}$ serves as a simulation parameter; we assume that the light source has a center wavelength $\lambda_c$ of 700 nm and the uniform spectral distribution from $\lambda_c- \frac{\Delta\lambda}{2}$ to $\lambda_c+\frac{\Delta\lambda}{2}$, where $\Delta\lambda$ denotes a wavelength width.  
    \item We use the stellar angular diameter as a simulation parameter; we assume the uniform intensity distribution of the light source within its angular diameter. 
    \item We take the planetary $x$- and $y$-directional separation angles $(\theta_x,\theta_y)$ (normalized by $\lambda_{c}/D$) as simulation parameters.
\end{enumerate}
{For the numerical Fourier transform, we use the same calculation method as the one shown in Appendix B of the paper \citet{2022AJ....163..279I}.}
\subsection{result}
\begin{figure}[htbp]
    \centering
    \includegraphics[width=0.9\textwidth]{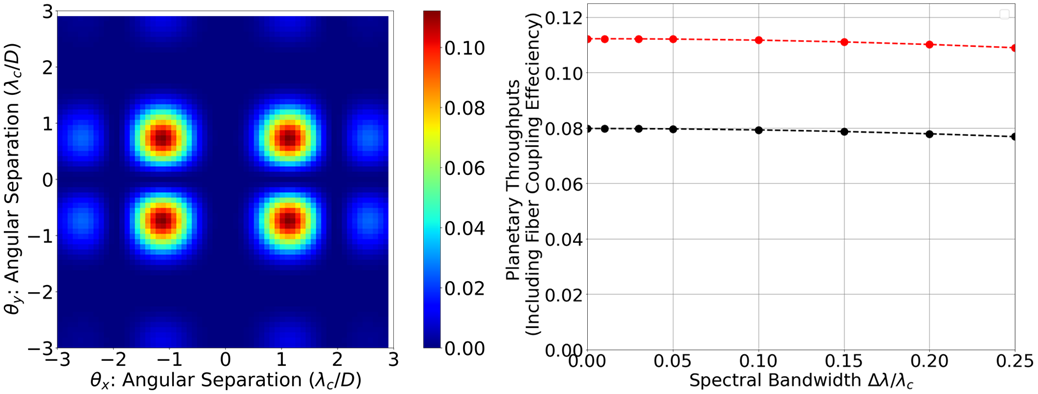}
    \caption{The colormap in the left panel shows planetary throughputs (including fiber-coupling efficiencies) in the case of different angular separations; the horizontal and vertical axes indicate the $x$- and $y$-directional separation angles normalized by $\lambda_{c}/D$, respectively. The calculation for the left panel assumes a monochromatic light source ($\Delta \lambda/\lambda_c=0$). The horizontal axis of the right panel indicates the spectral bandwidth $\Delta\lambda/\lambda_c$. The vertical axis of the right panel denotes the planetary throughput including the fiber-coupling efficiency. The markers in the right panel show the simulation results; the red and black markers indicate the case when the planetary separation angles $(\theta_x,\theta_y)=(1.125,0.750)$ and $(\theta_x,\theta_y)=(1.000,1.000)$, respectively. The dashed straight lines connect the markers as interpolation. }
    \label{fig:my_label3}
\end{figure}

Figure \ref{fig:my_label3} shows the simulation result concerning the planetary throughput.
The left panel of Figure \ref{fig:my_label3} indicates the planetary throughputs (including the fiber-coupling efficiency) over different planetary separations {assuming monochromatic light.}
Our original design philosophy of the Lyot-plane mask expected that the throughput map of the left panel of Figure \ref{fig:my_label3} has peak values when 
\begin{eqnarray}
    \left(\theta_x,\theta_y\right)&=&(q,r), (-q,r), (q,-r), (-q,-r) \nonumber\\
    &=& (1,1), (-1,1), (1,-1), (-1,-1).
\end{eqnarray}
However, the result in the left panel of Figure \ref{fig:my_label3} shows that the throughput map peaks when
\begin{eqnarray}
    \left(\theta_x,\theta_y\right)&=&(Q,R), (-Q,R), (Q,-R), (-Q,-R),
\end{eqnarray}
where $(Q,R)\sim(1.125, 0.750)$; the peak value reaches about 11\%.
The right panel of Figure \ref{fig:my_label3} shows the planetary throughputs for different spectral bandwidths. 
We can observe the constancy of the planetary throughputs against the change in the spectral bandwidth.

\begin{figure}[htbp]
    \centering
    \includegraphics[width=0.5\textwidth]{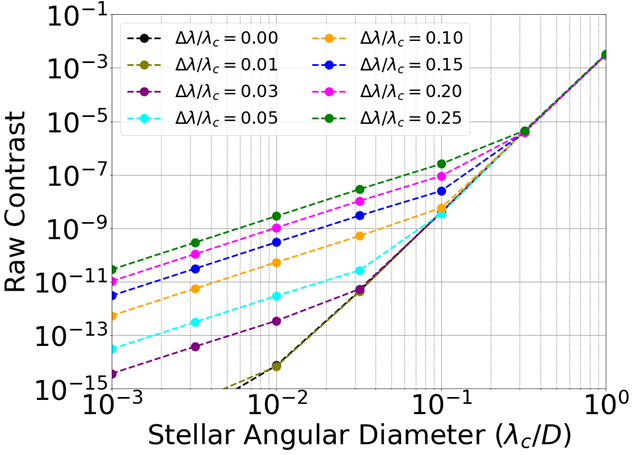}
    \caption{The vertical axis indicates the ratio (raw contrast) of the stellar and planetary throughputs including the fiber-coupling efficiency. The horizontal axis means the stellar angular diameter normalized by $\lambda_{c}/D$. The markers indicate the simulation results with the assumption that $(\theta_x,\theta_y)=(1.125,0.750)$. The colors of markers mean the difference in spectral bandwidths (see the legend). The dashed straight lines connect the markers as interpolation.}
    \label{fig:my_label4}
\end{figure}

Figure \ref{fig:my_label4} shows the simulation result concerning the achievable raw contrast.
In Figure \ref{fig:my_label4}, we can find that the new method can achieve the raw contrast of about $4\times10^{-8}$ ($\Delta \lambda/\lambda_c=0.25$) and $5\times10^{-10}$ ($\Delta \lambda/\lambda_c=0.10$) when the star has the angular diameter of $3\times 10^{-2}\ \lambda_{c}/D$.
We can also see that in the monochromatic case ($\Delta \lambda/\lambda_c=0.00$), the raw contrast seems proportional to the about sixth power of the stellar angular diameter.
In the cases with non-zero spectral bandwidths, the raw contrast and the stellar diameters exhibit no simple power-law relations, but when we focus only on the region of the stellar angular diameters less than $3\times 10^{-2}\ \lambda_{c}/D$, the relation approximately obeys power laws with the power law exponent of about two and the constant coefficients depending on the spectral bandwidths. 
On the other hand, when the stellar angular diameters overvalue $3\times 10^{-1}\ \lambda_{c}/D$, the raw contrasts have little dependency on the spectral bandwidth.
When we focus only on the intermediate region, we can infer that, in this region, the relations have different power law exponents depending on the spectral bandwidths. 
\section{Discussion} \label{sec:discussion}
\subsection{Planetary Throughput}
\subsubsection{Factor-Specific Discussion on the Planetary Throughput}
{We can interpret that the simulated peak value 11\% of the planetary throughput (Figure \ref{fig:my_label3}) comes from the multiplication of simulated fiber-coupling efficiency and} the following factors inevitable in the present method. (i)Because we use the 1DDLC, the throughput estimation must include a factor of the square of the mask constant $a$ (see Section \ref{suma}); $a^2=48.6$\%.
(ii)The Lyot-plane phase mask (two-dimensional diffraction grating) forms four major diffraction peaks of the planetary light with the same intensities on the focal plane. 
Nevertheless, only one of the peaks transmits through the single-mode fiber (Figure \ref{fig:my_label1}).
This effect, therefore, degrades the throughput by a factor of four (25\%).
Multiplying the factors (i) and (ii) leads to a value of 12.1\%.
{Compared with the value 12.1\%, the simulation result of the planetary throughput (including the fiber-coupling efficiency) of 11\% infers a high fiber-coupling efficiency of 91\%.}

\subsubsection{Azimuthal-Angle Dependency of Planetary Throughput}
\begin{figure}[htbp]
    \centering
    \includegraphics[width=1.0\textwidth]{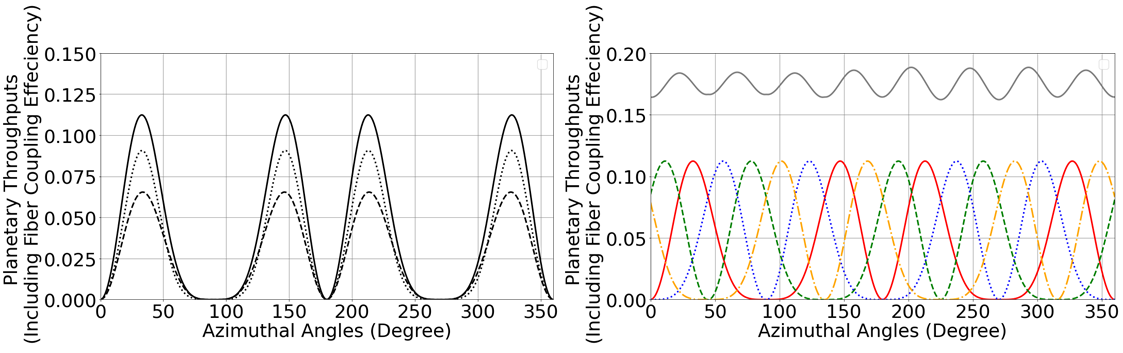}
    \caption{The vertical and horizontal axes mean the planetary throughput (including fiber coupling efficiency) and the azimuthal angles of the planetary separation, respectively. (Left) The dashed, solid, and dotted curves correspond to the cases of 1.000, 1.352 (the radius of the maximum throughputs), and 1.600 $\lambda_{c}/D$ of the radius of the planetary separation, respectively. {(Right) The red solid curve indicates the same curve as the solid curve in the left panel. The green dashed, blue dotted, and orange dash-dotted curves show the same ones shifted (rotated) along the horizontal axis by 45$^\circ$, 90$^\circ$, and 135$^\circ$, respectively. The gray solid curve denotes the sum of all four curves.  } }
    \label{fig:my_label4.5}
\end{figure}

In Figure \ref{fig:my_label3}, we observe that the planetary throughput of this method has a non-uniform dependency on the azimuthal angle for a given radius of the planetary separation.
We extracted and plotted the azimuthal-angle dependence of the planetary throughput in Figure \ref{fig:my_label4.5}. 
Here, we discuss the impact of the non-uniform throughput on planetary detection and spectroscopic observation.
\paragraph{Planetary Detection} 
Without anticipation of the planetary location, the azimuth dependence degrades the exploration efficiency (the left panel of Figure \ref{fig:my_label4.5}).
However, from a practical point of view, how much additional time we need is a more important measure than the degradation rate of the exploration efficiency.
In the right panel of Figure \ref{fig:my_label4.5}, we show that we can acquire an almost azimuthally homogeneous throughput with a throughput of about 18\% by summing the four exposures with the image rotation by 0$^{\circ}$, 45$^{\circ}$, 90$^{\circ}$, and 135$^{\circ}$.
Based on this scanning strategy, we can estimate the increase of the observation time $\Delta t_{p}$ required for $\mathrm{S/N}=p$ in the following manner:
\begin{equation}
    \Delta t_{p}=(Q-1)t_{p},
\end{equation}
where the degradation rate $Q$ of the observation efficiency satisfies $Q=4$ (the number of exposures required for the complete azimuthal scan) and $t_p$ denotes the observation time required for $\mathrm{S/N}=p$ in the case with no azimuth dependence of the throughput ($Q=1$).
Using the effective throughput $\tau$ ($\sim0.18$ from the right panel of Figure \ref{fig:my_label4.5}) and the number of planetary photons that reach the telescope aperture per hour $\Dot{n}$, we can express the requirement time $t_{p}$ in the case with no degradation in the observation efficiency as follows:
\begin{equation}
    t_{p}=\frac{p^2}{\tau\Dot{n}},
\end{equation}
where we ignored the photons of the stellar leak because the number of them falls less compared to the planetary photons thanks to the nuller's contrast-mitigation ability presented in Figure \ref{fig:my_label4}. 
In Table \ref{tab:my_label}, we show per-an-hour numbers $\Dot{n}$ of planetary photons that reach inside the telescope aperture for the possible target examples under a given condition ($D=6\ \mathrm{m}$, $\lambda_c=700\ \mathrm{nm}$, and $\Delta \lambda /\lambda_{c}=0.10$) assuming detecting potentially habitable planets around different types of host star. Using the values $\Dot{n}$ in Table \ref{tab:my_label}, we can estimate the increase of the observation time $\Delta t_{5}$ required for $\mathrm{S/N}=5$ as follows:
\begin{eqnarray}
    \Delta t_{5}&=&\begin{cases}
0.28\mathrm{h} \ \ \left(T_{\mathrm{eff}}=3900K\right) \\
0.32\mathrm{h}  \ \ \left(T_{\mathrm{eff}}=3500K\right).
\end{cases}
\end{eqnarray}
{Hence, even though exposure time quadruples, for typical target system of the 1DDLC, the overall detection time to achieve that $\mathrm{S/N} = 5$ remains under an hour.}
To change the planetary azimuthal angle with respect to the coronagraphic instruments, for example, we can rotate the entire telescope around the optical axis \citep{2020arXiv200106683G} or use an image-rotation unit made with a Dove prism \citep{2003OptCo.220..257M}.
\begin{table}[]
    \centering
    \begin{tabular}{cccc|cccc}
    \multicolumn{4}{c}{Assumed Parameters}&\multicolumn{4}{c}{Calculated Values}\\ \hline \hline
$T_{\mathrm{eff}}$ [K]&$d$ [pc]&$\theta_{\ast}$ [mas]&$\theta_{\ast}$ [$\lambda_{c}/D$]&$\theta_{\mathrm{sep}}$ [mas]&$\theta_{\mathrm{sep}}$ [$\lambda_{c}/D$]&$C$&$\Dot{n}$ [$\mathrm{h^{-1}}$]
\\ \hline
3900&10&0.56&2.3$\times10^{-2}$&27&1.1&2.9$\times10^{-9}$&1500\\ 
3500&10&0.36&1.6$\times10^{-2}$&15&0.62&1.0$\times10^{-8}$&1300\\ 
    \end{tabular}
    \caption{{Typical observation requirement for detecting potentially habitable planets around different types of host stars. The effective temperature $T_{\mathrm{eff}}$, distance $d$, and stellar angular diameter $\theta_{\ast}$ characterize the assumption of the host stars. The first, second, and third column in the table means the cases of an early-M dwarf ($T_{\mathrm{eff}}=3900K$) and mid-M dwarf ($T_{\mathrm{eff}}=3500K$), respectively. We referred to the values of the stellar physical radii $R_{\ast}$ in \citet{2012ApJ...757..112B} to assume the values of $\theta_{\ast}$. As values determining instrumental requirements, we show the planetary angular separation $\theta_{\mathrm{sep}}$, planet-star contrast ratio $C$, and the number of photons that reach the telescope aperture per hour after reflected at the planetary surfaces $\Dot{n}$ ($D=6\mathrm{m}$, $\lambda_c=700\mathrm{nm}$, and $\Delta \lambda /\lambda_{c}=0.10$) in the table. To evaluate $\theta_{\mathrm{sep}}$, we used the following equation determining the planet-star physical distance $r_{p}$ such that the planet receives the same amount of radiation flux as the Earth: $r_{p}=\mathrm{(1\ au)}\times(R_{\ast}/R_{\odot})(T_{\mathrm{eff}}/T_{\odot})^2$. The contrast $C$ comes from the following equation assuming Earth-sized face-on planets: $C=A_{g} (R_{\oplus}/r_{p})^2/\pi$, where $A_{g}$ denotes the geometric albedo of the Earth. We derived $\Dot{n}$ by multiplying the contrast $C$ by the number of stellar photons that reach the telescope aperture per hour evaluated with Planck's law. }}
    \label{tab:my_label}
\end{table}
\paragraph{Possibility of a Calibration Method for Planetary Detection}
The azimuth dependence of the planetary throughput may work as a key to further contrast mitigation through data analyses.
The curves in Figure \ref{fig:my_label4.5} show that the planetary signals have their inherent profile as functions of the azimuthal angles of the planetary separation.
On the other hand, the stellar leak is constant for the change of the planetary azimuthal angle in principle. 
Hence, for example, we can expect that using the cross-correlation method between the model profile of the signal and measured dataset along the azimuthal angles will lead to higher signal-to-noise ratios compared to ones expected from the raw contrast of Figure \ref{fig:my_label4}.
A similar data filtering technique appears in observing mid-infrared thermal emissions emitted from Earth-like planets with nulling interferometer \citep{1997Icar..128..202M}, but the present one and nulling interferometers have a difference in the point that the present one will remove stellar leaks (speckle noises) while the nulling interferometers will remove exozodiacal emissions. 
\paragraph{Spectroscopic Observation}
After detecting the location of a promising target, we can use a fixed image-rotation angle optimized to the target to perform the spectroscopic observation with the highest throughput (11\%).  
Since we need no image rotation during this observation mode, we can secure sufficient detected-photon number for each resolvable spectrum element with a practical integration time.

\subsection{Raw Contrast as a Function of Stellar Angular Diameter}
\subsubsection{Mathematical Interpretation}
\begin{figure}[htbp]
    \centering
    \includegraphics[width=0.4\textwidth]{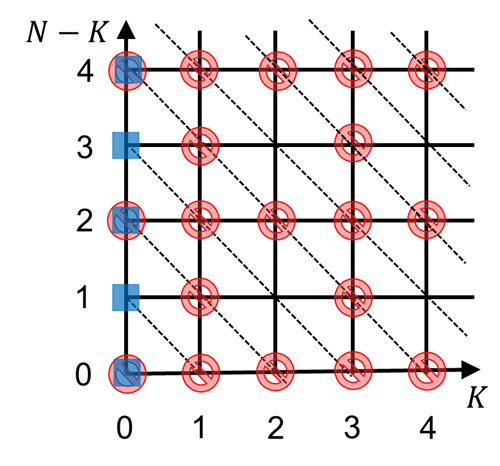}
    \caption{A schematic showing which terms of $\left\lbrace g_{K;N-K}(\alpha,\beta)| K,N \in \mathbb{Z} \right\rbrace$ can transmit the optical system of the new method. The horizontal and vertical axis indicates the values of $K$ and $N-K$, respectively. Every lattice point on the figure corresponds to a pair of integers $K$ and $N-K$. The diagonal dashed lines connect the lattice points belonging to the same values of $N$. The red prohibited symbols mean that the associated terms fail to transmit the single-mode fiber. The blue-filled squares denote the terms to be erased by the coronagraph system perfectly in the case of the zero spectral bandwidth.}
    \label{fig:my_label5}
\end{figure}

The numerical simulation suggests that the raw contrast approximately obeys a linear combination of some power-law functions consistently to the following mathematical interpretation.
Using Taylor expanding for functions with two variables, we can rewrite Equation (\ref{a9}) as the following expression:
\begin{equation}
v_{\mathrm{tilt}}\left(\vec{\alpha}\right)=P(\vec{\alpha})\sum_{N=0}^{\infty}\sum_{K=0}^{N}g_{K;N-K}(\alpha,\beta),
\end{equation}
where
\begin{equation}
g_{K;N-K}(\alpha,\beta)=\frac{(2\pi i)^N}{N!}\binom{N}{K}\left(\theta_x \alpha\right)^{K}\left(\theta_y \beta\right)^{N-K}.
\end{equation}
Since the function $g_{K;N-K}(\alpha,\beta)$ is a power function, the parity of the function $g_{K;N-K}(\alpha,\beta)$ concerning the arguments $\alpha$ and $\beta$ match the parity as integers of power exponents $K$ and $N-K$, respectively. 
Thus, only when $K$ is even and $N-K$ is odd, the term $g_{K;N-K}(\alpha,\beta)$ transmits the optical system of the new method (Figure \ref{1} and \ref{fig:my_label5}).
As reviewed in Section \ref{suma}, the 1DDLC erases the terms where $K=0$ perfectly when the spectral bandwidth has no width (Figure \ref{fig:my_label5}).
Conversely, in the cases with non-zero spectral bandwidths, the terms where $K=0$ can produce the leak amplitudes depending on the spectral bandwidths; we show the terms where $K=0$ with blue-filled squares in Figure \ref{fig:my_label5}.
From Figure \ref{fig:my_label5}, we can observe that, in the case with zero spectrum bandwidth, the term where $(K,N-K)=(2,1)$ corresponds to the least-order term ($N=3$) that can transmit the system.
We can also find that, in the case with non-zero spectrum bandwidth, the term where $(K,N-K)=(0,1)$ emerges as the least-order term ($N=1$) that can transmit the system.
Hence, because a term where $N=d$ can contribute to the $2d$-order sensitivity in the intensity leak, we can infer that the raw contrast $C(\Theta)$ as the function of the stellar angular diameter $\Theta$ approximately obeys the following functional form:
\begin{equation}
    C(\Theta)= c_2 \Theta^2 + c_6 \Theta^6,\label{fitfunc}
\end{equation}
where the symbols $c_2$ and $c_6$ denote constants; we can expect that only the constant $c_2$ depends on the spectral bandwidths from the above discussion.

\subsubsection{Fitting with a Parametric Curve}
\begin{figure}[htbp]
    \centering
    \includegraphics[width=1.0\textwidth]{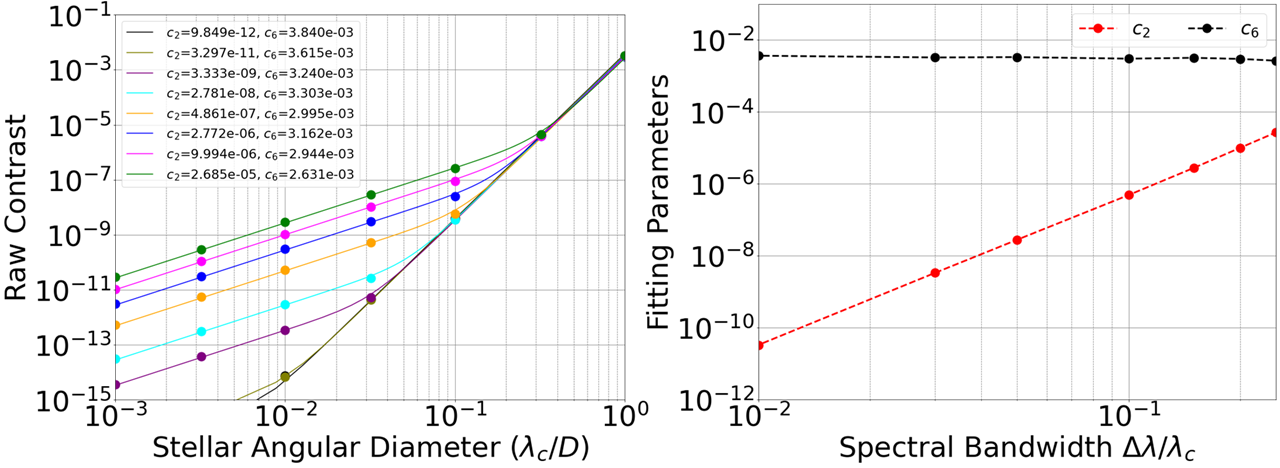}
    \caption{The markers in the left panel are the same as the ones in Figure \ref{fig:my_label4}. The curves in the left panel show the fitted function for the different spectral bandwidths.  The fitting assumes the functional form of Equation (\ref{fitfunc}). The legend in the left panel includes the fitting results of the fitting parameters $c_{2}$ and $c_{6}$.  The vertical axis of the right panel means the fitting results of the fitting parameters $c_{2}$ (red markers) and $c_{6}$ (black markers), and the horizontal axis denotes the spectral bandwidth. The dashed straight lines connect the markers as interpolation. }
    \label{fig:my_label6}
\end{figure}

We fit the functional form of Equation (\ref{fitfunc}) to the simulation results in Figure \ref{fig:my_label4}, adopting the weighted least-square method with the weight that is inversely proportional to the raw contrast data themselves.
The curves in the left panel of Figure \ref{fig:my_label6} show the values of the fitting functions to exhibit how much the fitting functions fit the dataset.
We plot the resultant values of the fitting parameters as functions of the spectral bandwidth in the right panel of Figure \ref{fig:my_label6}.  
In the right panel of Figure \ref{fig:my_label6}, we can find that the parameter $c_2$ is approximately proportional to the fourth power of the spectral bandwidth. In contrast, the parameter $c_6$ is practically independent of the spectral bandwidth.  

\section{Conclusion} \label{sec:conclusion}
In this study, we presented a new method to achieve a wideband nulling with immunity to stellar angular diameters by adding a Lyot-plane phase mask and a single-mode fiber located at the optical axis to a 1DDLC.
After explaining the theory of the new method, we conducted a numerical simulation of the method to know the theoretically achievable performance of the method.
The main findings from the simulation result include the following.
(i)When we treat the planetary throughput (including the fiber-coupling efficiency) as a function of the angular separation $\left(\theta_x,\theta_y\right)$ of the planets, the peak value of the function reaches about 11\%.
(ii)The peak values of the planetary throughput are almost independent of the spectral bandwidth. 
(iii)The new method can achieve the raw contrast of about $4\times10^{-8}$ (in the case with the spectral bandwidth of 25\%) and $5\times10^{-10}$ (the spectral bandwidth of 10\%) when the star has the angular diameter of $3\times 10^{-2}\ \lambda_{c}/D$.
A factor-specific discussion on the planetary throughput suggested a high fiber coupling efficiency of the present method.
We discussed the planetary-azimuthal-angle dependency of the planetary throughput. 
The dependence may degrade the exploration efficiency but, thanks to the wideband photometry, will make no unacceptable increase in the observation time required for the detection of the main targets of the present method.
We also mentioned the possibility of using the dependence for a calibration method. 
We have also discussed the mathematical interpretation of the simulation result and assumed a functional form to explain the simulation result.
The result of the functional fitting of the simulation results with the assumed function has suggested the following.
(1)We can consider the raw contrast as a function of the stellar diameter as a linear combination of the second-order and sixth-order power functions (Equation \ref{fitfunc}) with sufficient accuracy. 
(2)The weight factor of the linear combination for the second-power term is practically proportional to the fourth power of the spectral bandwidth, while the one for the sixth-power is approximately independent of the spectral bandwidth.

{The authors deeply appreciate the anonymous reviewers' contribution to improving the manuscript of this paper through their important comments. }
The authors are sincerely grateful to Dr. Takahiro Sumi a professor at Osaka University for encouraging our activity in this study.  

\appendix

\section{Another Expression of the Lyot-plane Mask \label{ape2}}
Using the Fourier series expansion, we can rewrite the Equation (\ref{eq9}) as the following:
\begin{eqnarray}
 s^{\mathrm{rect}}_{qr}\left(\vec{\alpha}\right)&=&\sum_{m=1}^{\infty}w_m\cos{\left(2\pi q(2m-1) \alpha\right)}\sum_{l=1}^{\infty}u_l\sin{\left(2\pi r(2l-1) \beta\right)}, \label{eq9_1}
\end{eqnarray}
where $w_m$ and $u_l$ denote the weight factors of the superposition of the sinusoidal functions; $w_m=\frac{4(-1)^{m-1}}{\pi(2m-1)}$ and $u_l=\frac{4}{\pi(2l-1)}$.  
The square moduli of weight factors $w_m$ and $u_l$ are proportional to the diffraction efficiency of the relevant diffraction order.

\bibliography{sample631.bbl}{}
\bibliographystyle{aasjournal}

\end{document}